\documentclass[aps,prl,twocolumn,footinbib]{revtex4-1}
\usepackage{graphicx}
\usepackage{indentfirst}
\usepackage{braket}
\usepackage{float}
\usepackage{amsmath}
\usepackage{epstopdf}
\usepackage{CJK}
\usepackage{esint}
\usepackage{color}
\usepackage[T1]{fontenc}
\usepackage{subfigure}
\usepackage{amsfonts}
\usepackage{footmisc}
\usepackage{scrextend}
\usepackage{multirow}
\usepackage{cleveref}
\usepackage[english]{babel}
\usepackage{url}

\def\be{\begin{equation}}
\def\ee{\end{equation}}
\def\ba{\begin{eqnarray}}
\def\ea{\end{eqnarray}}
\usepackage{bm}

\newcommand{\bt}{\boldsymbol \theta}

\begin{document}

\title{Entanglement-Enhanced Lidars for Simultaneous Range and Velocity Measurements }

\author{Quntao Zhuang$^{1,2}$}
\email{quntao@mit.edu}
\author{Zheshen Zhang$^1$}
\author{Jeffrey H. Shapiro$^1$}
\affiliation{$^1$Research Laboratory of Electronics, Massachusetts Institute of Technology, Cambridge, Massachusetts 02139, USA\\
$^2$Department of Physics, Massachusetts Institute of Technology, Cambridge, Massachusetts 02139, USA}
\date{\today}
\begin{abstract}
Lidar is a well known optical technology for measuring a target's range and radial velocity.  We describe two lidar systems that use entanglement between transmitted signals and retained idlers to obtain significant quantum enhancements in simultaneous measurement of these parameters.  The first entanglement-enhanced lidar circumvents the Arthurs-Kelly uncertainty relation for simultaneous measurement of range and radial velocity from detection of a single photon returned from the target.  This performance presumes there is no extraneous (background) light, but is robust to the roundtrip loss incurred by the signal photons.   The second entanglement-enhanced lidar---which requires a lossless, noiseless environment---realizes Heisenberg-limited accuracies for both its range and radial-velocity measurements, i.e., their root-mean-square estimation errors are both proportional to $1/M$ when $M$ signal photons are transmitted.  These two lidars derive their entanglement-based enhancements from use of a unitary transformation that takes a signal-idler photon pair with frequencies $\omega_S$ and $\omega_I$ and converts it to a signal-idler photon pair whose frequencies are $(\omega_S + \omega_I)/2$ and $\omega_S-\omega_I$.  Insight into how this transformation provides its benefits is provided through an analogy to superdense coding.
\end{abstract}

\maketitle

Quantum metrology~\cite{Helstrom_1976,giovannetti2006,giovannetti2011advances} addresses measuring unknown parameters of a physical system using quantum-mechanical resources.  A typical single-parameter scenario involves interrogating a physical system with $M$ probes that undergo independent, identical interactions with the system.  These probes then carry away information about the unknown parameter of interest that can be used to estimate its value.  When the $M$ probes are in a product state, the standard quantum limit (SQL)---with root-mean-square (rms) estimation error proportional to $1/\sqrt{M}$---can be achieved.  Entangled probes, however, offer a quantum enhancement in our single-parameter setting~\cite{giovannetti2006,giovannetti2011advances} that can realize the Heisenberg limit (HL), viz., an rms estimation error that is proportional to $1/M$~\cite{zwierz2010general,Giovannetti_2001,giovannetti2004,giovannetti2006,giovannetti2011advances,escher2011}. 
The preceding SQL versus HL behavior for single-parameter estimation arises in many interesting scenarios, e.g., in measuring time delays~\cite{Giovannetti_2001}, point-source separations~\cite{nair2016,lupo2016,Tsang_2016,kerviche2017}, displacements~\cite{Genoni_2013,steinlechner2013quantum,ast2016reduction}, or magnetic fields~\cite{Baumgratz_2016}.

Significant complications occur, in the independent, identical interactions setting, when multiple unknown parameters are to be estimated~\cite{Genoni_2013,steinlechner2013quantum,ast2016reduction,Baumgratz_2016}.  In particular, if these parameters are associated with noncommuting observables, then the uncertainty principle would seem to forbid obtaining unlimited simultaneous knowledge of them from a single returned probe~\cite{Fujikawa_2014,Ozawa_2003,Fujikawa_2012,Heisenberg_1927,Kennard_1927,Robertson_1929,Arthurs_1965}. 
In such cases quantum-enhanced accuracy can be obtained by entangling probes with locally-stored idlers~\cite{ballester2004entanglement,fischer2001enhanced,fujiwara2001quantum,fujiwara2003quantum,ballester2004estimation,Genoni_2013,steinlechner2013quantum,ast2016reduction}, in addition to the benefit derived from entangling different probes.  Moreover, 
the optimum probe state can be entangled in a complicated form, and the optimum measurement scheme is collective~\cite{ballester2004estimation}, thus making both difficult to determine~\cite{Fujiwara_2001}, even when the observables for the parameters of interest commute~\cite{Humphreys_2013}. 

In this Letter we address a specific instance of quantum metrology for a pair of parameters that are associated with noncommuting observables:  the lidar problem of measuring both a target's range and its radial velocity.  Specifically, we describe two lidar systems that use entanglement between transmitted signals and retained idlers to obtain significant quantum enhancements in simultaneous measurement of these parameters.  The first entanglement-enhanced lidar circumvents the Arthurs-Kelly uncertainty relation~\cite{Arthurs_1965} for simultaneous measurement of range and radial velocity from detection of a single photon returned from the target.  
This performance presumes there is no extraneous (background) light, but is robust to the roundtrip loss incurred by the signal photons.  For comparison, a system that does not use entanglement would need to detect two returned signal photons to achieve the same measurement performance.  Thus our system's advantage can be quite significant when the lidar-to-target-to-lidar path is very lossy.  Note that it had previously been thought~\cite{Yuen_2009,Nair_2011} that there was no entanglement advantage to be had in lossy, noiseless lidar scenarios~\cite{Lloyd2008,Tan2008,Guha2009,Zheshen_15,Zhuang2017}.

Our second entanglement-enhanced lidar---which requires a lossless, noiseless environment---realizes HL accuracies for both its range and radial-velocity measurements, i.e., their root-mean-square estimation errors are both proportional to $1/M$ when $M$ signal photons are transmitted.  For comparison, both the $M$-photon time-domain and $M$-photon frequency-domain Giovannetti-Lloyd-Maccone (GLM) states~\cite{Giovannetti_2001} must probe the target if the same performance as our system's is to be obtained when there are no retained idlers.  Thus our system's advantage can be quite significant when the probing flux must be kept as low as possible to prevent damage to delicate targets, e.g., in sensing the microscopic motions of biological samples.  

Both of the preceding lidars derive their entanglement-enhanced performance from use of a unitary transformation that takes a signal-idler photon pair with frequencies $\omega_S$ and $\omega_I$ and converts it to a signal-idler photon pair whose frequencies are $(\omega_S + \omega_I)/2$ and $\omega_S-\omega_I$.  Interestingly, as we will show, using this transformation makes our entanglement-enhanced lidars behave in a continuous-parameter manner that is analogous to the discrete-parameter behavior of superdense coding (SDC)~\cite{bennett1992,Mozes_2005} in quantum communication.   

\begin{figure}
\centering
\includegraphics[width=0.5\textwidth]{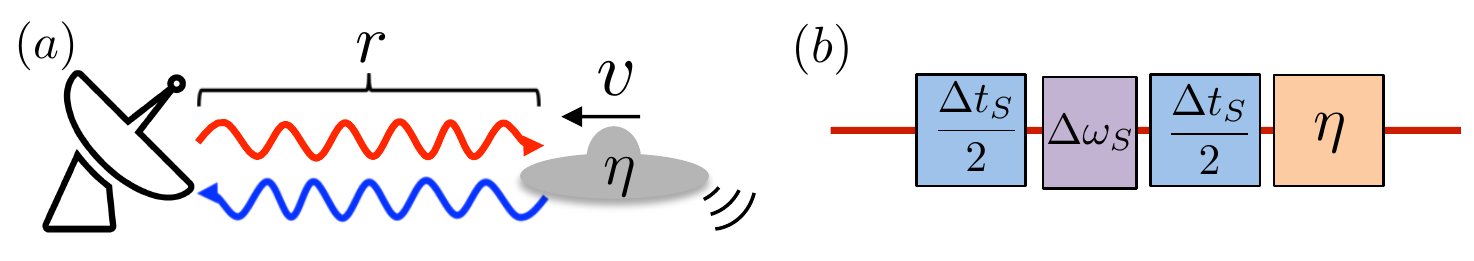}
\caption{(a) Lidar sensing of target range and radial velocity. (b) Equivalent quantum-channel representation.
}
\label{Fig1main}
\end{figure}

{\em Lidar Range and Radial-Velocity Estimation.---}
Consider the lidar sensing problem shown in Fig.~\ref{Fig1main}(a).  A collection of $M$ quasimonochromatic signal photons with center frequency $\omega_{S_c}$ are directed toward a target whose range, $r$, and radial velocity, $v$ (with $v>0$ indicating a target moving toward the lidar), are to be estimated from the time delay, $\Delta t_S = 2r/c$, and the Doppler shift, $\Delta \omega_S = 2\omega_{S_c} v/c$, that are imposed on each photon that returns to the lidar, where $c$ is light speed.  

For a lidar that performs single-mode detection at its receiver, background light at optical wavelengths of interest can be small enough to be ignored, e.g., background light will have an average of $\sim$$10^{-6}$ photons per mode in daytime operation at the 1.55\,$\mu$m eyesafe wavelength~\cite{SGE2005}. Thus, aside from the time delay and Doppler shift incurred by each photon, the only propagation effect we shall consider for the lidar-to-target-to-lidar channel is its roundtrip transmissivity, $\eta$, which will typically satisfy $\eta \ll 1$, making photon efficiency a priority in estimating a target's range and its radial velocity from the $\eta M$ photons, on average, that are returned.  

Figure~\ref{Fig1main}(b) shows an equivalent channel model for the Fig.~\ref{Fig1main}(a) configuration.  Each signal photon incurs a time delay $\Delta t_S$/2 on its way to the target, a Doppler shift $\Delta \omega_S$ upon reflection from the target, another $\Delta t_S/2$ time delay en route back to the lidar, where (without loss of generality) we have chosen to impose the roundtrip transmissivity $\eta$.  For convenience, in what follows, we will use $\hat{D}_{S_t}(\Delta t_S/2)$ and to denote the operator that time delays a signal photon by $\Delta t_S/2$, and $\hat{D}_{S_\omega}(\Delta \omega_S)$ to denote the operator that Doppler shifts a signal photon by $\Delta \omega_S$.  

We will begin our development of entanglement-enhanced lidar sensing by considering the best that can be done when only one photon is returned from the target to the lidar.  Suppose that $M$ transmitted photons are emitted one at a time by the lidar's transmitter, and that we know both those emission times \emph{and} which transmitted photon resulted in the one returned to the lidar.  So, because a target's range and its radial velocity are then easily calculated from that photon's time of arrival and its Doppler shift, all that follows will address limits of simultaneous time and frequency measurements.  Furthermore, in our quest for quantum enhancement, we will assume that each signal photon is entangled with a retained idler photon in an initial pure state $|\psi\rangle$ and that each idler is stored, in a lossless manner, for a time $\Delta t_I$ that is sufficient to enable its being jointly measured with its signal-photon companion should that companion be the one that is returned to the lidar.

{\em Cram\'{e}r-Rao bound for pure-state observations.---}
The Cram\'{e}r-Rao (CR) bound~\cite{Helstrom_1976,Holevo_1982,Yuen_1973} sets the ultimate limit on the accuracy with which our lidar can estimate the arrival time and Doppler shift of its single returned photon based on a joint measurement of that photon together with its retained idler companion. Let $|\psi(\bt)\rangle$ be the joint state of these returned and retained photons, where $\bt = [\Delta t_S,\Delta \omega_S]^T$ with $^T$ denoting transpose, and let the positive operator-valued measurement $\hat{\Pi}_{\tilde{\bt}}$ be an unbiased estimator of $\bt$, i.e, $\int\!{\rm d}\tilde{\bt}\,\tilde{\bt}\,\langle\psi(\bt)|\hat{\Pi}_{\tilde{\bt}}|\psi(\bt)\rangle = \bt$.  The CR bound for this case sets the following lower limit~\cite{Fujiwara_1994} on this estimator's error-covariance matrix, ${\bf V}(\bt) = \langle (\tilde{\bt} - \bt) (\tilde{\bt} - \bt)^T\rangle$:  
\begin{eqnarray}
\lefteqn{{\rm tr}[{\bf G}{\bf V}(\bt)] \ge {\rm tr}\!\left[{\bf G}{\bf J}_{\bt}^{-1}\right]} \nonumber \\
&& +\,\,\frac{\sqrt{{\rm det}({\bf G})}}{{\rm det}({\bf J}_{\bt})} |\braket{\psi(\bt)|[L_{\Delta t_S},L_{\Delta\omega_S}]|\psi(\bt)}|.
\label{CRpure_main}
\end{eqnarray}
In this inequality:  ${\bf G}$ is an arbitrary $2\times 2$ positive-semidefinite real-valued cost matrix; ${\det}(\cdot)$ denotes determinant; $[\hat{a},\hat{b}]\equiv \hat{a}\hat{b}-\hat{b}\hat{a}$ is the commutator $\hat{a}$ and $\hat{b}$;  $\hat{J}_{\bt}$ is the quantum Fisher-information matrix, whose $jk$th element, for $j,k = \Delta t_S, \Delta\omega_S$, is 
$({\bf J}_{\bt})_{jk}\equiv4 \left[{\rm Re}\left( \partial_{\theta_j}\bra{\psi(\bt)}\partial_{\theta_k}\ket{\psi(\bt)}\right)\right.$ $\left. + \bra{\psi(\bt)} \partial_{\theta_j} \ket{\psi(\bt)}\bra{\psi(\bt)} \partial_{\theta_k} \ket{\psi(\bt)}\right]
$; and $L_j \equiv2\left(\partial_{\theta_j}\ket{\psi(\bt)} \bra{\psi(\bt)}+ \ket{\psi(\bt)} \partial_{\theta_j}\bra{\psi(\bt)}\right)$, for $j = \Delta t_S, \Delta\omega_S$, are the symmetric logarithmic derivatives.

{\em Single-photon target-return lidar.---}
When the lidar-to-target-to-lidar roundtrip transmissivity is very low, i.e., $\eta \ll 1$, then transmission of $M \simeq 1/\eta \gg 1$ photons is necessary for a reasonable assurance that one signal photon will be returned from that target to the lidar's receiver.   To minimize the $M$ value needed to estimate target range and radial velocity it would be best were it possible to simultaneously---and accurately---determine the time delay $\Delta t_S$ and the Doppler shift $\Delta\omega_S$ from measurement of a single returned photon.   
This wish would seem to violate the Arthurs-Kelly uncertainty relation~\cite{Arthurs_1965}, which states that $\delta t_S$ and $\delta \omega_S$---the rms errors when time delay and Doppler shift are estimated from such a simultaneous measurement---satisfy
$\delta t_S \, \delta \omega_S\ge 1.$  However, because our lidar has the retained idler photon for use in a joint measurement with its returned-signal companion, we will see that the Arthurs-Kelly inequality can be circumvented.  Indeed, starting from a biphoton state with time-bandwidth product $TW\gg 1$, we will show how $\delta t_S  \simeq 1/2W$ and $\delta \omega_S \simeq 1/2T$ can be achieved simultaneously from an appropriate joint measurement.

Our single-photon lidar will use a nondegenerate spontaneous parametric downconverter (SPDC) whose output---for the signal-idler pair that will ultimately be measured---can be taken to be the biphoton state $\ket{\psi}=\int\!{\rm d}t_S\, {\rm d}t_I\, \psi(t_S,t_I) \ket{t_S}_S\ket{t_I}_I$ with time-domain wave function given by~\cite{Zhang_2014,supp}
\be
\psi(t_S,t_I) \propto e^{-t_-^2/4\sigma_{\rm cor}^2-t_+^2/4\sigma_{\rm coh}^2-i(\Delta \omega t_-/2 +\omega_Pt_+)},
\label{wavefunc_time_omega_main}
\ee
where $|t\rangle$ denotes a single photon at time $t$, $t_-\equiv t_S-t_I$, $t_+ \equiv (t_S+t_I)/2$, $\sigma_{\rm cor}$ is the biphoton correlation time, $\sigma_{\rm coh}$ is the pump coherence time, $\Delta\omega \equiv \omega_{S_c}-\omega_{I_c}$ is the difference between the signal and idler's center frequencies, and $\omega_P$ is the pump frequency. 
This state's frequency-domain representation, $\int\!{\rm d}\omega_S\, {\rm d}\omega_I\, \Psi(\omega_S,\omega_I) \ket{\omega_S}_S\ket{\omega_I}_I$ where $|\omega\rangle$ denotes a single photon with frequency $\omega$, then has wave function 
\be
\Psi(\omega_S,\omega_I)\propto e^{-(\omega_--\Delta\omega)^2\sigma_{\rm cor}^2/4-(2\omega_+-\omega_P)^2\sigma_{\rm coh}^2},
\ee
with $\omega_-\equiv \omega_S-\omega_I$ and $\omega_+ \equiv (\omega_S+\omega_I)/2$.

The rms time durations of the the SPDC's signal and idler photons are identical, and given by $T = \sqrt{\sigma^2_{\rm coh} + \sigma^2_{\rm cor}/4}$. Likewise, their rms bandwidths are also identical, and given by $W = \sqrt{1/16\sigma^2_{\rm coh} + 1/4\sigma^2_{\rm cor}}$.   When $\sigma_{\rm cor} = 2\sigma_{\rm coh}$, the biphoton reduces to a product of pure-state signal and idler photons satisfying $TW =1/2$.  A continuous-wave downconverter, however, typically has $\sigma_{\rm coh} \gg \sigma_{\rm cor}$, so that $T\approx \sigma_{\rm coh} \gg 1/W \approx 2\sigma_{\rm cor}$, making the signal and idler highly entangled, with entanglement entropy $S_E=\log_2\left(2TW\right) \gg 1$. 

Conditioned on the biphoton from Eq.~(\ref{wavefunc_time_omega_main}) being the one whose returned signal and retained idler will be measured, we have that $|\psi(\bt)\rangle = \hat{D}_{S_t}(\Delta t_S/2)\hat{D}_{S_\omega}(\Delta\omega_S)\hat{D}_{S_t}(\Delta t_S/2)\otimes \hat{D}_{I_t}(\Delta t_I)|\psi\rangle$ is the state from which we will determine the signal photon's time delay and Doppler shift.
Using the CR bound (\ref{CRpure_main}), we can show~\cite{supp} that unbiased estimators of the signal photon's time delay and Doppler shift have rms errors that individually obey $\delta t_S\ge 1/2W$ and $\delta \omega_S \ge 1/2T$, and jointly satisfy
\be 
\delta t_S \, \delta \omega_S\ge {\left(1+2TW\right)}/\left({8 T^2W^2}\right).
\ee 
Without entanglement ($TW = 1/2$), we recover the Arthurs-Kelly inequality, but with highly-entangled signal and idler ($TW\gg 1$), we get $\delta t_S\,\delta \omega_S \ge 1/4TW$, which suggests that $\delta t_S = 1/2W$ and $\delta \omega_S = 1/2T$ might be realized simultaneously.  We next present a theoretical design for achieving that goal.    

Our first step is to apply the single-photon unitary transformation~\cite{causality}
\ba
\hat{B}_{SI}&=& \int\!{\rm d}\omega_S\int\!{\rm d}\omega_I\,\left|\frac{\omega_S+\omega_I}{2}\right\rangle_S\ket{\omega_S-\omega_I}_I{}_S\!\bra{\omega_S}{}_I\!\bra{\omega_I}
\nonumber
\\
&=&\int\!{\rm d} t_{S}\int\!{\rm d} t_{I}\, 
\ket{t_{S}+t_{I}}_S\left|\frac{t_{S}-t_{I}}{2}\right\rangle_I {}_S\!\bra{t_{S}} {}_I\!\bra{t_{I}}.
\label{UX1}
\ea
to the postselected state $|\psi(\bt)\rangle$ to obtain the product state, $\hat{B}_{SI}|\psi(\bt)\rangle = |\psi_S(\bt)\rangle_S\otimes |\psi_I(\bt)\rangle_I$, where, assuming $\sigma_{\rm coh} \gg \sigma_{\rm cor}$, we have that
\begin{eqnarray}
\lefteqn{|\psi_S(\bt)\rangle_S \propto } \nonumber \\
&&\int\!{\rm d} \omega_S\,  e^{i(\omega_S+\Delta\omega)(\Delta t_S + \Delta t_I) -(2\omega_S-\omega_P-\Delta \omega_S)^2T^2 } \ket{\omega_S}_S, 
\end{eqnarray}
and
\be
|\psi_I(\bt)\rangle_I \propto \int\!{\rm d} t_I\, e^{-i(\Delta\omega_S+\Delta\omega)t_I-(2t_I-\Delta t_S+\Delta t_I)^2W^2} \ket{t_I}_I.
\ee
Next, we measure the single-photon frequency observable of the signal photon and the single-photon arrival-time observable of the idler photon, i.e., $\hat{\omega}_S = \int\!{\rm d}\omega_S\,\omega_S|\omega_S\rangle_S{}_S\langle \omega_S|$ and $\hat{t}_I = \int\!{\rm d}t_I\,t_I|t_I\rangle_I{}_I\langle t_I|$.  Using the resulting outcomes, $\tilde{\omega}_S$ and $\tilde{t}_I$,  we generate our time-delay and Doppler-shift estimates $\widetilde{\Delta t_S} = 2\tilde{t}_I + \Delta t_I$ and $\widetilde{\Delta\omega_S} = 2\tilde{\omega}_S - \omega_P$.  These estimates are unbiased, $\langle \widetilde{\Delta t_S} \rangle = \Delta t_S$ and $\langle \widetilde{\Delta\omega_S} \rangle = \Delta\omega_S$, with standard deviations
$1/2W$ and $1/2T$, thus showing that our entanglement-enhanced lidar \emph{simultaneously} realizes the CR bounds on $\delta t_S$ and $\delta \omega_S$ from a single-photon target return.  
\begin{figure}
\centering
\includegraphics[width=0.4\textwidth]{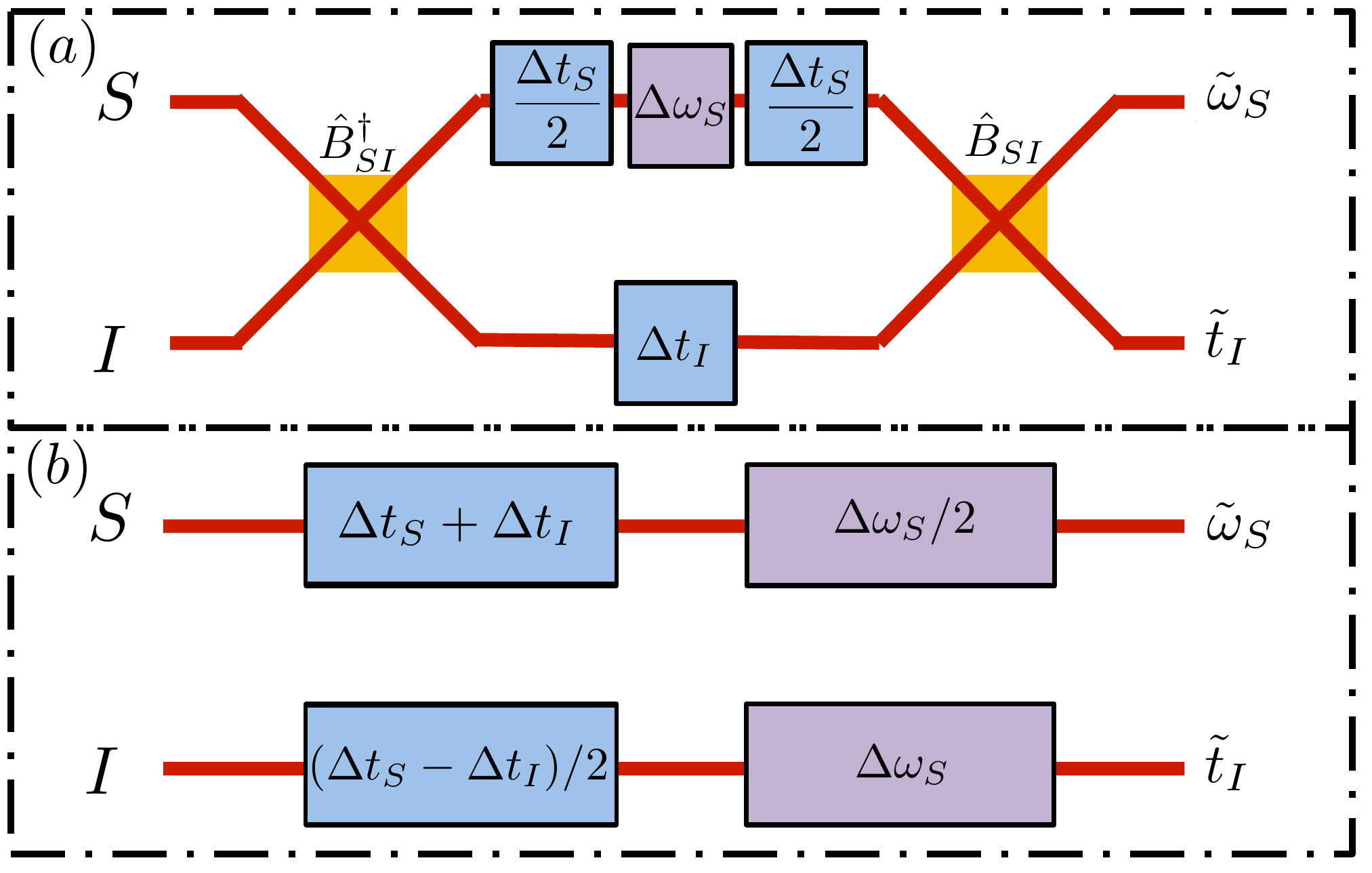}
\caption{(a) Entangled-state and (b) product-state representations of the single-photon target-return lidar.  See text for details.}
\label{Fig2main}
\end{figure}

{\em Connection to SDC.---} Figure~\ref{Fig2main} shows entangled-state and product-state representations of our single-photon lidar for the photon pair that is ultimately measured.  In Fig.~\ref{Fig2main}(a), we start from a signal-idler ($S$-$I$) product state whose frequency-domain wave function is proportional to $\exp(- 4\omega_S^2\sigma^2_{\rm coh}- \omega_I^2\sigma^2_{\rm cor}/4)$.  Applying the single-photon unitary transformation $\hat{B}_{SI}^\dagger$ to this state then 
yields the biphoton state from Eq.~(\ref{wavefunc_time_omega_main}).  After that biphoton's signal and idler undergo the time-delay and Doppler-shift transformations shown in Fig.~\ref{Fig2main}(a), application of $\hat{B}_{SI}$ converts them back to a product state, from which  a signal-photon frequency measurement, $\tilde{\omega}_S$, and an idler-photon arrival-time measurement, $\tilde{t}_I$, provide the information needed for simultaneous Doppler-shift and time-delay estimates.  The process---from product-state source output to product-state measurement input---is thus governed by the single-photon unitary transformation
\begin{align}
\hat{U}&\equiv
\hat{B}_{SI}[\hat{D}_{S_t}({\Delta t_S/2})
\hat{D}_{S_\omega}({\Delta \omega_S})
\hat{D}_{S_t}({\Delta t_S/2})\nonumber \\ 
&\otimes
\hat{D}_{I_t}(\Delta t_I)]\hat{B}_{SI}^\dagger.
\end{align}
After simple algebra, $\hat{U}$ can be rewritten---up to a global phase---as shown in Fig.~\ref{Fig2main}(b):  
\ba
\hat{U}=[\hat{D}_{S_t}(\Delta t_S+\Delta t_I)\hat{D}_{S_\omega}(\Delta \omega_S/2)]
\nonumber
\\
\otimes[\hat{D}_{I_t}((\Delta t_S-\Delta t_I)/2)\hat{D}_{I_\omega}(\Delta \omega_S)].
\label{Uequivalence} 
\ea
This form of $\hat{U}$ acting directly on the same signal-idler product state that was the input in Fig.~\ref{Fig2main}(a) immediately leads to our single-photon lidar's being able to sense Doppler shift from a signal-photon measurement and arrival time from an idler-photon measurement.  

The preceding representations of $\hat{U}$ comprise a continuous-variable analog of qubit SDC~\cite{bennett1992,Mozes_2005}, as we now show. In qubit SDC, Alice first applies a controlled-not (CNOT) gate to the state $\ket{\psi_0}_{AB}=\left(\ket{0}_A+\ket{1}_A\right)/\sqrt{2}\otimes \ket{0}_B$ to obtain a Bell state, and then sends the $B$ half to Bob as an ancilla.  Next, to send two classical bits---$b_1$ and $b_2$ = 0 or 1---to Bob on a single photon, Alice applies the Pauli operators $\hat{Z}_A^{b_2}\hat{X}_A^{b_1}$ to her half of the Bell state and transmits that encoded state to Bob.  Then, Bob applies a CNOT to the two photons in his possession, after which local measurements will suffice to recover Alice's bit values.  These three steps---entanglement preparation, encoding, and product-state recovery---are just like the $\hat{B}_{SI}^\dagger$, time delays and Doppler shift, and $\hat{B}_{SI}$  transformations shown in Fig.~\ref{Fig2main}(a).  Indeed, SDC's unitary transformation has the equivalent forms
$\hat{U}_{\rm SDC}={\rm CNOT}_{AB}(\hat{Z}_A^{b_2} \hat{X}_A^{b_1} \otimes \mathcal{I}_B){\rm CNOT}_{AB}
=(\hat{Z}_A^{b_2}\hat{X}_A^{b_1})\otimes \hat{X}_B^{b_1}$, which are discrete-variable versions of what we have seen in Figs.~\ref{Fig2main}(a) and (b), respectively.  

{\em Lidar with simultaneous HL scaling.---} 
\begin{figure}
\centering
\includegraphics[width=0.45\textwidth]{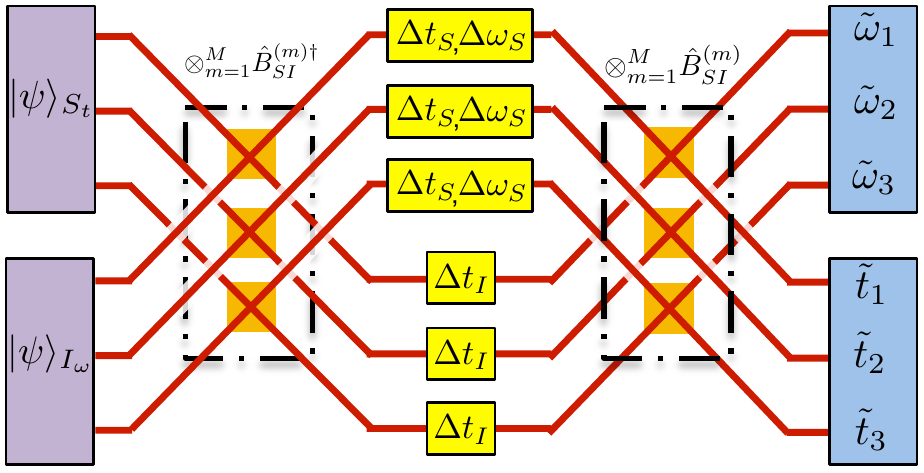}
\caption{Schematic for simultaneous time-delay and Doppler-shift measurements with HL rms accuracies ($M=3$).
\label{Fig3main}
}
\end{figure}
Giovannetti, Lloyd, and Maccone showed~\cite{Giovannetti_2001} that when the $M$-photon, $M$-mode, frequency-domain GLM signal state, $|\psi\rangle_{S_\omega} \propto \int\!{\rm d}\omega_S\,e^{-\omega_S^2/4W^2}\otimes_{m=1}^M|\omega_S\rangle_{S_m}$ interrogates a perfectly-reflecting target, then time-resolved detection of all $M$ photons in the absence of background noise enables $\Delta t_S$, the roundtrip time delay for that target, to be estimated with HL rms accuracy $\delta t_S = 1/2MW$.  Likewise, the $M$-photon, $M$-mode, time-domain GLM signal state, $|\psi\rangle_{S_t} \propto \int\!{\rm d}t_S\,e^{-t_S^2/4T^2}\otimes_{m=1}^M|t_S\rangle_{S_m}$, enables that target's Doppler shift, $\Delta \omega_S$, to be estimated with HL rms accuracy $\delta\omega_S = 1/2MW$ in this lossless and noiseless scenario.  These measurements, however, are an either-or proposition, viz., if an $M$-photon GLM state is used to interrogate the target we cannot get both $\delta t_S = 1/2MW$ and $\delta \omega_S = 1/2T$.  We can, by using an $M/2$-photon, frequency-domain GLM state, followed by an $M/2$-photon, time-domain GLM state, get time-delay and Doppler-shift measurements with rms accuracies $\delta t_S = 1/4MW$ and $\delta \omega_S = 1/4MT$.  Our second lidar will realize $\delta t_s  = 1/2MW$ and $\delta \omega_S = 1/2MT$ from transmission of $M$ signal photons toward the target.  

To simultaneously achieve HL accuracies, we employ two GLM states together with the $M$-mode generalization our first lidar's $\hat{U}$ transformation, see Fig.~\ref{Fig3main}.  We start from GLM signal and idler states $|\psi\rangle_{S_t}$ and $|\psi\rangle_{I_\omega}$ that are entangled by the application of $\otimes_{m=1}^M\hat{B}_{SI_m}^\dagger$.  Next, the signal photons illuminate and return from the target, having accumulated a roundtrip delay $\Delta t_S$ and a Doppler shift $\Delta\omega_S$, while the idler photons are stored at the lidar for a time $\Delta t_I$.  Applying $\otimes_{m=1}^M\hat{B}_{SI_m}$ to the returned and retained photons then undoes the entanglement.  Paralleling the development of Eq.~(\ref{Uequivalence}), we find that the overall state transformation accomplished by this arrangement is 
\ba
\hat{U}_M&=&\left[\otimes_{m=1}^M
\hat{D}_{{S_t}_m}(\Delta t_S+\Delta t_I)
\hat{D}_{{S_\omega}_m}(\Delta \omega_S/2)
\right]
\otimes
\nonumber
\\
&&\left[\otimes_{m=1}^M 
\hat{D}_{{I_t}_m}((\Delta t_S-\Delta t_I)/2) 
\hat{D}_{{I_\omega}_m}(\Delta \omega_S)\right].
\label{Uequivalence_M} 
\ea
It now follows immediately that a Doppler shift measurement on the $\hat{U}_M$-transformed signal photons has HL rms accuracy $\delta\omega_S = 1/2MT$ and a time-delay measurement on the $\hat{U}_M$-transformed idler photons has HL rms accuracy $1/2MT$.  

Note that GLM states are not normalizable.  So, to be more rigorous, we should redo the preceding development using a limiting procedure on the normalized GLM-like states that were introduced in Ref.~\cite{shapiro2007}.

{\em Discussion.---}
We have exhibited lidars that provide entanglement-enhanced accuracies in the simultaneous measurement of target range and radial velocity.  These parameters, which are linked to the time delay and Doppler shift of photons returning from the target, are associated with the noncommuting observables $\hat{t} = \int\!{\rm d}t\,t\,|t\rangle\langle t|$ and $\hat{\omega}= \int\!{\rm d}\omega\,\omega\,|\omega\rangle\langle\omega|$ for a single photon's arrival time and frequency.  Our general scheme of transforming operations with noncommuting generators to commuting observables can be applied to other simultaneous-measurement scenarios that involve noncommuting generators. Note that we use the number of probes (signal photons) in our resource counting, as has been done in prior work~\cite{zwierz2010general,giovannetti2011advances}.  References~\cite{Genoni_2013,steinlechner2013quantum,ast2016reduction} treat the $M=1$ case of simultaneous measurement of two quadratures, which is different from the multi-mode time-frequency measurement addressed in this Letter.  Means for implementing the $\hat{B}_{SI}$ transformation---using single-photon $\chi^{(2)}$ interactions and linear optics~\cite{Niu2017,NiuArXiv}---are under investigation.

\begin{acknowledgements}
This research was supported by Air Force Office of Scientific Research Grant No. FA9550-14-1-0052. Q. Z.  acknowledges support from the Claude E. Shannon Research Assistantship.
\end{acknowledgements}

\pagebreak
\begin{center}
\begin{widetext}
\textbf{\large Supplemental Material for: \\
``Entanglement-Enhanced Lidars for Simultaneous Range and Velocity Measurements''}
\end{widetext}
\end{center}
\setcounter{equation}{0}
\setcounter{figure}{0}
\setcounter{table}{0}
\setcounter{page}{1}
\makeatletter
\renewcommand{\theequation}{S\arabic{equation}}
\renewcommand{\thefigure}{S\arabic{figure}}
\renewcommand{\bibnumfmt}[1]{[S#1]}
\renewcommand{\citenumfont}[1]{S#1}
\begin{widetext}

\section{Single-photon time and frequency states and observables}
The states $|t\rangle$ and $|\omega\rangle$ represent single photons at time $t$ and frequency $\omega$, respectively.  They satisfy\begin{subequations}
\ba
&\braket{t|t^\prime}=\delta\left(t-t^\prime\right),
&\braket{\omega|\omega^\prime}=\delta\left(\omega-\omega^\prime\right),
\\
&\ket{\omega}= \int\!\frac{{\rm d}t}{\sqrt{2\pi}}\, e^{-i \omega t} \ket{t},\ \ \ \
&
\ket{t}=\int\!\frac{{\rm d}\omega}{\sqrt{2\pi}}\, e^{i \omega t} \ket{\omega},
\ea
\label{def}
\end{subequations}
and are the eigenkets of the single-photon time and frequency operators
\be
\hat{t} = \int\!{\rm d}t\,t\,|t\rangle\langle t|, \quad \hat{\omega} = \int\!{\rm d}\omega\,\omega\,|\omega\rangle\langle \omega|.
\ee
These operators have the nonzero commutator, 
\be
[\hat{\omega},\hat{t}] = i,
\ee
for which the Arthurs-Kelly uncertainty relation is
\be 
\delta t\,\delta \omega \ge 1.
\ee

\section{Time-energy entanglement of the Gaussian wave-function biphoton}
The biphoton state $|\psi\rangle$ from the main paper's Eq.~(2) has time-domain and frequency-domain wave functions given by 
\be
\psi(t_S,t_I) =\frac{\exp\!\left(-(t_S-t_I)^2/4\sigma_{\rm cor}^2-(t_S+t_I)^2/16 \sigma_{\rm coh}^2-i[\Delta\omega(t_S-t_I)+\omega_P(t_S+t_I)]/2\right)}{\sqrt{2\pi \sigma_{\rm coh}\sigma_{\rm cor}}} .
\label{wavefunc_t}
\ee
and
\be
\Psi(\omega_S,\omega_I)=\sqrt{\frac{2\sigma_{\rm coh}\sigma_{\rm cor}}{\pi}}\,
\exp\!\left(-\sigma_{\rm cor}^2\left(\omega_S-\omega_I-\Delta\omega\right)^2/4-\sigma_{\rm coh}^2\left(\omega_S+\omega_I-\omega_P\right)^2 \right).
\label{wavefunc_omega}
\ee

The signal photon's reduced density operator can be obtained by tracing out the idler's state,
\be
\hat{\rho}_S=\int\!{\rm d} t_I'\, {}_I\!\bra{t_I'}\ket{\psi}\bra{\psi}\ket{t_I'}_I,
\ee
from which we can obtain
\be
\delta t_S^2 \equiv {\rm tr}(\hat{\rho}_S\hat{t}_S^2) = \sigma^2_{\rm coh} + \sigma^2_{\rm cor}/4,
\ee
and
\be
\delta \omega_S^2 \equiv {\rm tr}[\hat{\rho}_S(\hat{\omega}_S-\omega_{S_c})^2] = 1/16\sigma^2_{\rm coh} + 1/4\sigma^2_{\rm cor}.
\ee
By symmetry, we have 
\be
\delta t_I^2 \equiv {\rm tr}(\hat{\rho}_I\hat{t}_I^2) = \delta t_S^2,
\ee 
and 
\be 
\delta \omega_I^2 \equiv {\rm tr}[\hat{\rho}_I(\hat{\omega}_I-\omega_{I_c})^2] = \delta \omega_S^2.
\ee

The entanglement entropy $S_E$ of the biphoton $|\psi\rangle$ equals the von Neumann entropy of $\hat{\rho}_S$.  The latter can be found, by analogy with the entropy calculation for the reduced density operator of a two-mode squeezed state, to be 
\be
S(\hat{\rho}_S) \equiv -{\rm tr}[\hat{\rho}_S\log_2(\hat{\rho}_S)] = \log_2(2\mu_A),
\ee
where
\be
\mu_A = \sqrt{(\sigma^2_{\rm coh}/\sigma^2_{\rm cor} + \sigma^2_{\rm cor}/16\sigma^2_{\rm coh})/4 + 1/8} = \log_2(2\delta t_S\delta\omega_S).
\ee
In the main paper we used this result with $T\equiv \delta t_S$ and $W \equiv \delta\omega_S$.

\section{Cram\'{e}r-Rao bound for time and frequency estimation}

The biphoton state at the single-photon lidar receiver in the main paper, $|\psi(\bt)\rangle$ is
\ba
\ket{\psi(\bt)}&=&
\int\!{\rm d}t_S \!\int\!{\rm d}t_I\,\psi(t_S-\Delta t_S,t_I-\Delta t_I) \exp[-i\Delta\omega_S(t_S-\Delta t_S/2)]\ket{t_S}_S\ket{t_I}_I
\\
&=& 
\int\!{\rm d} \omega_S\!\int\!{\rm d} \omega_I\, \Psi(\omega_S-\Delta \omega_S,\omega_I) \exp\left(i\Delta t_S\left(\omega_S-\Delta \omega_S/2\right) +i \omega_I \Delta t_I\right)\ket{\omega_S}_S\ket{\omega_I}_I.
\ea
It is now straightforward to obtain
\be
{\bf J}_{\bt} =  4\,{\rm diag}\left[W^2,T^2\right], 
\ee
and 
\be
|\langle \psi(\bt)|[L_{\Delta t_S},L_{\Delta \omega_S}]|\psi(\bt)\rangle|=4.
\ee
Using the main paper's (1) with $G={\rm diag}[1,0]$ and ${\rm diag}[0,1]$ then gives us $\delta t_S \ge 1/2W$ and $\delta \omega_S \ge 1/2T$.   Finally, using the main paper's (1) with $G = {\rm diag}[W^2, zT^2]$ and $z\ge 0$, we find that
\begin{align}
\delta t_S^2\, \delta \omega_S^2&\ge
\max_{z\ge0}
\left[\frac{\delta \omega_S^2}{W^2}\left(\frac{1}{4}+\frac{1}{4}\sqrt{\frac{z}{T^2W^2}}+z\left(\frac{1}{4}-T^2 \delta \omega_S^2\right)\right) \right]
\\
&=\frac{\delta \omega_S^2}{W^2} \left(\frac{1}{4}+\frac{1}{16T^2W^2\left(4T^2\delta \omega_S^2-1 \right)}\right)
\ge \frac{\left(1+2TW\right)^2}{64 T^4W^4},
\end{align}
which is the main paper's (3).

\end{widetext}


\begin{thebibliography}{99}
\bibitem{causality}For brevity, we have omitted the time delays that are needed to make this transformation causal.  Assuming a nondegenerate SPDC ensures that both post-$\hat{B}_{SI}$ photons will be at optical frequencies.
\end{thebibliography}
\end{document}